\documentclass{aa}  

\usepackage{graphicx}
\usepackage[varg]{txfonts}
\usepackage{soul}
\usepackage{xcolor}

\defcitealias{Sahade2020}{S20}

\begin{document}

   \title{Polarity relevance in flux-rope trajectory deflections \\ triggered by coronal holes}

   \author{A. Sahade
          \inst{1,2,3}
          \and
          M. C\'ecere\inst{1,3}
          \and
          A. Costa\inst{1}
          \and
          H. Cremades\inst{4}
          }

   \institute{Instituto de Astronom\'{\i}a Te\'orica y Experimental, CONICET-UNC, C\'ordoba, Argentina.\\
   \email{acosta@unc.edu.ar}
         \and
            Facultad de Matem\'atica, Astronom\'{\i}a, F\'{\i}sica y Computaci\'on, Universidad Nacional de C\'ordoba (UNC), C\'ordoba, Argentina.\\
            \email{asahade@unc.edu.ar}
         \and
             Observatorio Astron\'omico de C\'ordoba, UNC, C\'ordoba, Argentina.\\
             \email{mariana.cecere@unc.edu.ar}
        \and
             Universidad Tecnol\'ogica Nacional\,--\,Facultad Regional Mendoza, CONICET, CEDS, Rodriguez 243, Mendoza, Argentina.\\
             }

   \date{Received ; accepted }

 
  \abstract
    {Many observations suggest that coronal holes (CHs) are capable of deviating the trajectory of coronal mass ejections (CMEs) away from them. However, for some peculiar events, the eruption has been reported to be initially pulled towards the CH and then away from it. 
    }
    {We study the interaction between flux-ropes (FRs) and CHs by means of numerical simulations, with the ultimate goal of understanding how CHs can deviate erupting CMEs/FRs from purely radial trajectories.
    }
    {We perform 2.5D magnetohydrodynamical numerical simulations of FRs and CHs interacting under different relative polarity configurations. In addition, we reconstruct the 3D trajectory and magnetic environment of a particular event seen by the STEREO spacecraft on 30 April 2012, whose trajectory initially departed from the radial direction toward the CH but later moved away from it. 
    }
    {The numerical simulations indicate that at low coronal heights, depending on the relative magnetic field polarity between FR and CH, the initial deflection is attractive, i.e. the FR moves towards the CH (case of anti-aligned polarities) or repulsive, i.e. the FR moves away from the CH (case of aligned polarities). This is likely due to the formation of vanishing magnetic field regions or null points, located between the FR and CH (case of anti-aligned polarities) or at the other side of the FR (case of aligned polarities). The analysed observational event shows a double-deflection compatible with an anti-aligned configuration of magnetic polarities, which is supported by SDO observations. We successfully reproduce the double deflection of the observed event by means of a numerical simulation.
    
    }
   {}

   \maketitle
%

\section{Introduction}

Large amounts of mass and magnetic field detachments are involved in the release of coronal mass  ejections (CMEs). Many CMEs show clear indications of an embedded magnetic flux-rope structure (FR, i.e. magnetic field lines twisted around an axial field). The FR magnetic system, usually associated with filaments, frequently deflects from its outward radial direction. This may happen due to multiple factors. For example, intrinsic CME and filament properties may affect the final amount of deflection \citep[e.g.,][]{filippov2001, martin2003, panasenco2008, bemporad2009, panasenco2011, pevtsov2012, liewer2013,Mostl2015, Wang2015, Kay2015, Kay2017}.
The magnetic field environment also contributes to the deflection through the interaction of the CME/FR with surrounding structures, for example: coronal holes \citep[CH, e.g.,][]{Cremades2006,Xie2009,Gopalswamy2009,Kilpua2009,Panasenco2013SoPh}, active regions \citep[e.g.,][]{Kay2015,Mostl2015}, pseudostreamers \citep[e.g.,][]{Lynch2013ApJ}, streamer belts \citep[e.g.,][]{Zucarello2012,Kay2013,Yang2018}, and heliospheric current sheets \citep[e.g.,][]{Liewer2015}.

Previous reports have suggested that CME trajectories depend on the local and global gradients of the magnetic pressure \citep{Panasenco2013SoPh,Liewer2015, Sieyra2020}. They found that the analysed CMEs propagate in the direction of least resistance, away from CHs. This behaviour was also numerically studied by \citet[hereafter S20]{Sahade2020}, who found that the FR deflection away from the CH was firstly due to the minimum magnetic energy location  and secondly, due to the channelling imposed by magnetic field lines. 

Studies based on the evolution of CMEs in the high corona agree on that CHs and open magnetic fluxes, acting as strong ``magnetic walls'', repel neighbouring CMEs by deflecting their trajectories \citep{Cremades2006,Gopalswamy2009,Gui2011,Yang2018,Cecere2020}. However, there are few peculiar examples for which CME paths behave differently. \cite{Jiang2007ApJ} investigated a CME that first evolved toward a CH and later moved away from it.  Moreover, \citet[][see their figure 14]{Sieyra2020} found that the trajectories of some of the analysed prominence-CMEs were not always deflected away from CHs. Particularly at low coronal heights, the eruption occasionally approached the CH, even when the trajectory was not directed toward the region of minimum magnetic energy.  \cite{Yang2018} studied the interaction of a small filament with non-CH open field lines. The filament polarity was anti-aligned with respect to the polarity of the open magnetic field lines, meaning that the positive polarity footpoint of the filament was adjacent to the negative polarity of the open magnetic fluxes. 
They found that the erupted filament material first approached the open field lines and afterwards was strongly deflected away from them. 

To improve understanding of the interaction between CMEs and CHs we perform 2.5D MHD numerical simulations of a FR in the low corona. 
We analyse the resulting trajectory deflection  and we study  the initial forces acting on the FR (Section 2). We consider both, aligned and anti-aligned polarities of the CH open field lines with respect to the nearest FR footpoint polarity. In Section 3 we study an eruptive event that interacted with a CH, observed on 30 April 2012 by the \textit{Solar-Terrestrial Relations Observatory} \citep[STEREO,][]{kaiser2008} twin spacecraft. It initially approached the CH and then propagated away from it, similarly as the peculiar observed eruptive events mentioned above.
Conclusions and some open questions are presented in Section 4.

\section{Numerical simulation}

\begin{figure}
    \centering
    \includegraphics[width=0.4\textwidth]{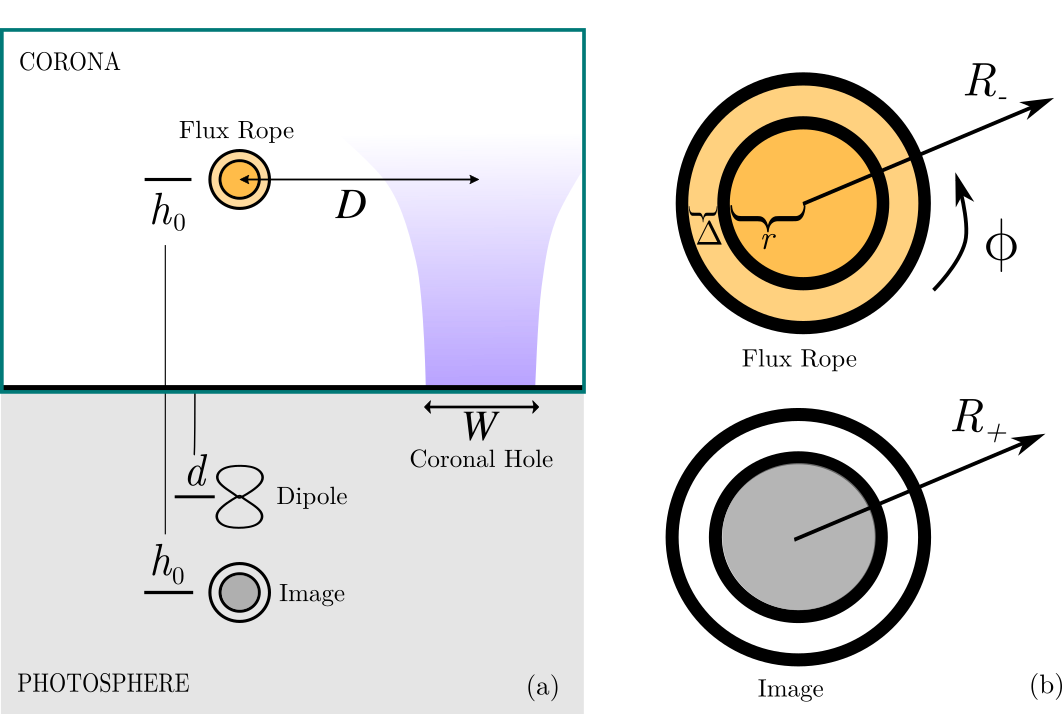}
    \caption{(a) Flux-rope and coronal hole scheme  not to scale. The turquoise frame indicates the simulated region. $h_0$ is the  FR (orange circle)  height and $D$ is the FR distance to the CH (purple-shaded area) whose width is parametrised by $W$. The line dipole and the image current are located at depth $d$ and $h_0$, respectively. (b) Current wire layers: $r$ is the radius, $\Delta$ is the thickness of the transition layer, $R_{-}$ is the radial coordinate from the FR centre,$R_{+}$ is the radial coordinate from the image wire , and $\phi$ the poloidal coordinate. }
    \label{fig:my_label}
\end{figure}

We perform numerical simulations to study the interaction between a FR and a CH without other interfering magnetic structures. We use a simple 2.5 dimensional model that provides meaningful information about magnetic fields and FR dynamics. The interaction is solved through the ideal MHD equations in presence of a gravitational field. The equations in CGS units in the Cartesian conservative form are written as:

\begin{equation}\label{e:cont}
\frac{\partial\rho}{\partial t}+\nabla\cdot(\rho\vec{v})=0 \, ,
\end{equation}
\begin{equation}\label{e:euler}
\frac{\partial (\rho \vec{v})}{\partial t} + \nabla \cdot \left(\rho \vec{v} \vec{v} - \frac{1}{4\pi} \vec{B}\vec{B} \right) + \nabla p + \nabla\left( \frac{B^2}{8\pi}\right)   = \rho \vec{g} \, ,
\end{equation}
\begin{equation}\label{e:consE}
\frac{\partial E}{\partial t} + \nabla \cdot \left[\left(E + p + \frac{B^2}{8\pi}\right)\vec{v} -\frac{1}{4\pi} \left(\vec{v\cdot B}\right)\vec{B}\right] = \rho \vec{g v} \, ,
\end{equation}
\begin{equation}\label{e:induccion}
\frac{\partial \vec{B}}{\partial t} + \vec{\nabla \cdot} \left(\vec{v} \vec{B} - \vec{B} \vec{v} \right) = \vec{0} \, ,
\end{equation}

\noindent
where $\rho$ represents the plasma density, $p$ the thermal pressure, $\vec{v}$ the velocity, $\vec{B}$ the magnetic field, and $\vec{g}$ the gravity acceleration. $E$ is the total energy (per unit volume), given by  
\begin{equation*}
    E = \rho \epsilon + \frac{1}{2} \rho v^2 + \frac{B^2}{8\pi},
\end{equation*}
where $\epsilon$ is the internal energy and
\begin{equation*}
{\vec{j}=\frac{c}{4\pi}{\nabla\times}\vec{B}, } \,
\end{equation*}
is the current density, with $c$ being the speed of light.

In addition to the MHD equations, the divergence-free condition of the magnetic field must be fulfilled, i.e.
\begin{equation}\label{e:divB}
 \vec{\nabla\cdot} \vec{B} = 0\, .
\end{equation}

For completeness we assume a perfect gas for which $p = 2\rho k_B T/m_i = (\gamma - 1) \rho \epsilon$, where $k_B$ is the Boltzmann constant, $T$ the plasma temperature, $m_i$ the proton mass (assuming that the medium is a fully ionised hydrogen plasma), and $\gamma = 5/3$ the specific heat relation. 

Simulations were performed using the FLASH Code \citep{2000ApJS..131..273F} in its fourth version, operated with the USM (unsplit staggered mesh) solver, which uses a second-order directionally unsplit scheme with a MUSCL-type reconstruction.
Outflow conditions (zero-gradient) are used at lateral and upper boundaries, line-tied condition is used at lower boundary. To preserve the initial force-free configuration outside the FR, a linear extrapolation of the magnetic field is established (Zurbriggen,  private communication). We can neglect the magnetic resistivity and use the ideal MHD equations. This results in a significant reduction of the computational cost, since the numerical diffusion present in the simulations provides the necessary dissipation \citep{Krause2018}.
The highest  resolution corresponds to $\sim[0.1 \times 0.1]~\textrm{Mm}^2$ cells, in a $[-700,700]~\textrm{Mm}\times [0,700]~\textrm{Mm}$ physical domain, where pressure and temperature gradients satisfy the refinement criterion.

\subsection{Magnetic model}
We depart from an out-of-equilibrium  magnetic FR, whose general configuration is schematised in Fig.~\ref{fig:my_label}.
The magnetic field of the FR is produced by a current wire, an image current wire and a line dipole. The image current is located below the photosphere having an  opposite direction, so as to generate a repulsive force. The line dipole, located below the photosphere,  provides an attractive force to the CME wire and emulates the photospheric field. The equations are piecewise-defined in three zones:
\begin{description}
    \item[Z1] Inside a current wire, $0\leq R < r-\frac{\Delta}{2}$.
    \item [Z2] Throughout the transition layer, $r-\frac{\Delta}{2}\!\leq \!R\! <r+\frac{\Delta}{2}$.
    \item [Z3] Outside a current wire, $r+\frac{\Delta}{2} \leq R$,
\end{description}
where $r$ is the current wire radius, $\Delta$ is the thickness of the transition layer between the current wire and the exterior, and $R$ is the radial coordinate from the centre of the current wire (Fig.~\ref{fig:my_label}b).

The magnetic field component $B_\phi$ generated by a current wire with current distribution $j_z$ is given by: 

\begin{equation} \label{e:Bphi}
   \!\!\!\!\!\! B_\phi(R)\!=\!
    \left\{
\begin{array}{rl}
\begin{alignedat}{2}
&\tfrac{2\pi}{c}j_0R  &&\text{\small at Z1}\\
&\tfrac{2\pi j_0}{cR}\left\{\tfrac{1}{2}\left(r-\tfrac{\Delta}{2}\right)^2-\left(\tfrac{\Delta}{2}\right)^2+\right. 
\\
&\tfrac{R^2}{2}+\tfrac{\Delta R}{\pi}\text{sin}\left[\tfrac{\pi}{\Delta}\left(R-r+\tfrac{\Delta}{2}\right)\right]+

\\
&\left.\!\!\!\left(\tfrac{\Delta }{\pi}\right)^2\cos\left[\tfrac{\pi}{\Delta}\left(R-r+\tfrac{\Delta}{2}\right)\right]\right\} \qquad && \text{\small at Z2}\\
&\tfrac{2\pi j_0}{cR}\left[r^2+\left(\tfrac{\Delta}{2}\right)^2-2\left(\tfrac{\Delta}{\pi}\right)^2\right]  && \text{\small at Z3,}
\end{alignedat}
\end{array} \right.
\end{equation}
\begin{equation} \label{e:jz}
   \! j_\mathrm{z}(R)\!=\!
    \left\{
\begin{array}{rl}
\begin{alignedat}{2}
&\!j_0 &&\text{\small at Z1}\\
&\!\tfrac{j_0}{2}\left\{\cos\left[\tfrac{\pi}{\Delta}\left(R-r+\tfrac{\Delta}{2}\right)\right]+1\right\} \quad  && \text{\small at Z2}\\
&\! 0  && \text{\small at Z3;}
\end{alignedat}
\end{array} \right.
\end{equation}
\noindent
where $j_0$ is a current density. 

In order to obtain a helical magnetic field in the FR we include a magnetic field in the $z$-axis of strength $B_\mathrm{z}$. In this way, as we showed in \citetalias{Sahade2020}, we avoid excessive gas pressure values needed to balance the magnetic pressure inside the flux-rope in the initial equilibrium state. The component $B_\mathrm{z}$ of the magnetic field and the current distribution $j_{\phi}$, are described by:
\begin{equation}\label{e:Bfield}
    B_\mathrm{z}(R) = \tfrac{\sqrt{8}\pi j_1}{c}\sqrt{\left(r-\tfrac{\Delta}{2}\right)^2-R^2}\, , 
\end{equation}
\begin{equation}\label{e:jphi}
    j_\phi(R) = j_1R\left[\sqrt{\left(r-\tfrac{\Delta}{2}\right)^2-R^2}\right]^{-1}\,, 
\end{equation}
where $j_1$ is a current density. These expressions are valid inside the flux-rope (Z1) and are null in the rest of the domain. 

For the initial magnetic field of the CH we use the same expressions as in \citetalias{Sahade2020}. The total initial magnetic field is the sum of the magnetic field of the FR and the magnetic field of  the CH, which in Cartesian components is given by:
\begin{align}\label{e:BfieldFR}
    B_x=&B_0\ \sin\left(\frac{x-D}{W}\right)\, \exp[-y/W]-B_\phi(R_-)\tfrac{(y-h_0)}{R_-} +  \nonumber \\
    & \: B_\phi(R_+)\tfrac{(y+h_0)}{R_+} -MdB_\phi{\scriptstyle\left(r+\tfrac{\Delta}{2}\right)}\left(r+\tfrac{\Delta}{2}\right)\tfrac{x^2-(y+d)^2}{R_d^4} \, ,\nonumber \\
    B_y= &B_0\ \cos\left(\frac{x-D}{W}\right)\,\exp[-y/W] + B_\phi(R_-)\tfrac{x}{R_-} -  \nonumber \\
    & B_\phi(R_+)\tfrac{x}{R_+}-MdB_\phi{\scriptstyle\left(r+\tfrac{\Delta}{2}\right)}\left(r+\tfrac{\Delta}{2}\right)\tfrac{2x(y+d)}{R_d^4}  \, , \nonumber \\
    B_z=&B_{\text{z}}(R_-)\, .
\end{align}
\noindent
where the parameter $B_0$ is the radial magnetic field strength of the CH at the distance $D$ on the $x$-axis. The parameter $W$ is related to the width of the CH (see next subsection) and modifies the decay of the magnetic field strength in the $y$-direction. $h_0$ is the initial vertical position of the FR and $M$ is the intensity of the line dipole at depth $d$. 
The distances $R$ are:
\begin{equation*}
    R_\pm = \sqrt{x^2+(y\pm h_0)^2},
\end{equation*}
\begin{equation*}
    R_d = \sqrt{x^2+(y+d)^2},
\end{equation*}
where $R_{-}, R_{+}$ and $R_{d}$ originate in the FR centre, image current wire and line dipole, respectively (see Fig. 1 of \citetalias{Sahade2020} for more details).

\subsection{Thermodynamic variables}

We simulate the solar atmosphere by adopting a multi-layer structure \citep{Mei2012}. The chromosphere lies between $y=0$ and $y=h_\mathrm{ch}$ with constant temperature $T_\mathrm{ch}$. The transition region, located between $y=h_\mathrm{ch}$ and the base of the corona ($y=h_\mathrm{c}$), is represented by a linearly increasing temperature up to $T_\mathrm{c}$, which is the constant temperature assumed for the corona. Thereby, the initial temperature distribution is given by
\begin{equation}
    {\textstyle T(y)=}
    \left\{
\begin{array}{rl}
\begin{alignedat}{2}
&{\textstyle T_\mathrm{ch}} &&\text{\small if  $0\leq y < h_\mathrm{ch}$}\\
&{\textstyle (T_\mathrm{c}-T_\mathrm{ch})\left[\frac{y-h_\mathrm{ch}}{h_\mathrm{c}-h_\mathrm{ch}}\right]+T_\mathrm{ch}} \quad && \text{\small if  $h_\mathrm{ch}\leq y < h_\mathrm{c}$}\\
& {\textstyle T_\mathrm{c}}  && \text{\small if  $h_\mathrm{c} \leq y$ }.
\end{alignedat}
\end{array} \right.  
\end{equation}
We set a temperature of $T_\mathrm{ch}=10\,000\,\text{K}$ for the chromosphere and $T_\mathrm{c}=10^6\,\text{K}$ for the corona. The height of the chromosphere is  $h_\mathrm{ch}=10\,\text{Mm}$, while the base of the corona is at $h_\mathrm{c}=15\,\text{Mm}$.

The temperature inside the FR ($T_{\text{\tiny{FR}}}$) varies according to the following temperature distribution:
\begin{equation}
    \!T(R_-)\!=\!
    \left\{
\begin{array}{rl}
\begin{alignedat}{2}
&\!T_{\text{\tiny{FR}}} && \text{\small at Z1}\\
&\!\!(T_\mathrm{c}\!-\!T_{\text{\tiny{FR}}})\!\left[\tfrac{R_--(r+\Delta/2)}{\Delta}\right]\!+\!T_{\text{\tiny{FR}}} \quad && \text{\small at Z2}\\
&\! T_\mathrm{c} && \text{\small at Z3.}
\end{alignedat}
\end{array} \right.
\end{equation}

In the previous work \citepalias{Sahade2020}, we considered the current-free atmosphere in hydrostatic equilibrium. Hence, the pressure $p(y)$ was only a function of $y$ considering a system having the $y$-axis aligned to the gravity acceleration  (i.e., $\vec{g} =\frac{-G M_\sun}{(y + R_\sun)^2}\vec{e}_y$, where $G$ is the gravitational constant, $M_\sun$ is the Sun mass, $R_\sun$ is the solar radius, and $y = 0$ corresponds to the solar surface). In this work we add a dimensionless factor $\chi$ to the pressure distribution, to set a subdense CH \citep[it is based on the density distributions proposed by][]{Pascoe2014}

\begin{equation} \label{subdens}
   \! \chi(x,y)\!=\!
    \left\{
\begin{array}{rl}
\begin{alignedat}{2}
&\!\beta(x,y)+1 &&\text{\small if $y<W/2$}\\
&\!\left(2-2\frac{y}{W}\right)\,\beta(x,y)+1 \quad  && \text{\small if $W/2 \leq y<W$}\\
&\! 1  && \text{\small if $W\leq y$,}
\end{alignedat}
\end{array} \right.
\end{equation}
where
\begin{equation}
    \beta(x,y) =\left(\frac{n_\mathrm{CH}}{n_\mathrm{c}} -1\right)\mathrm{sech}^2\left(\frac{x-D}{W\arcsin{(0.25 \exp[y/W}])}\right)^2 \,. \nonumber
\end{equation}
The number density at height $y = h_\mathrm{c}$ in the corona is $n_\mathrm{c}$, while inside the CH the number density decays to  $n_\mathrm{CH}$. According to the resulting subdense zone, the effective width of the CH is set to $w \sim W/2 $.
Then the pressure is $P(x,y) = \chi(x,y) p(y)$, where $p(y)$ is the hydrostatic component:
 
\begin{equation}\label{pres}
    {\textstyle  p(y)=\!}
    \left\{
\begin{array}{rl}
\begin{alignedat}{2}
&{\textstyle \!\!p_\mathrm{ch}\exp{\! \left[\frac{\alpha}{T_\mathrm{ch}}\left(\frac{1}{h_\mathrm{ch}+R_{\sun}}-\frac{1}{y+R_{\sun}}\right)\right]}} &&\text{\small if  $0\leq y < h_\mathrm{ch}$}\\
&{\textstyle\! \!p_\mathrm{ch}\exp{\!\left[-\int_{h_\mathrm{ch}}^{y}\frac{\alpha}{T{\scriptstyle(y')}}(R_{\sun}+y')^{-2} dy'\right]}} \quad && \text{\small if  $h_\mathrm{ch}\leq y < h_\mathrm{c}$}\\
&{\textstyle \! \!\frac{k_B}{N_Am_i}T_\mathrm{c}n_\mathrm{c}\exp{\!\left[-\frac{\alpha}{T_\mathrm{c}}\left(\frac{1}{h_\mathrm{c}+R_{\sun}}-\frac{1}{y+R_{\sun}}\right)\right]}}  && \text{\small if  $h_\mathrm{c} \leq y$,}

\end{alignedat}
\end{array} \right.  
\end{equation}
where 
\begin{equation*}
    {\textstyle p_\mathrm{ch}(y)=\frac{k_B}{N_Am_i}T_\mathrm{c}n_\mathrm{c}\exp{\left[\int_{h_\mathrm{ch}}^{h_\mathrm{c}}\frac{\alpha}{T(y')}(R_{\sun}+y')^{-2} dy'\right]}} \, ,
\end{equation*}
$\alpha= \frac{m_i G M_\sun}{2k_B}$, and $N_A$ is the Avogadro number.
The internal pressure of the FR is obtained by proposing a solution close to the equilibrium:
\begin{align} \label{e:presion}
    p_{\text{\tiny{FR}}}(x,y) = P(x,y)& +\tfrac{1}{c}\int_{R}^{r+\frac{\Delta}{2}}B_\phi{\scriptstyle(R')}j_z{\scriptstyle(R')}dR'\nonumber\\
    &-\tfrac{1}{c}\int_{R}^{r+\frac{\Delta}{2}}B_{\text{z}}{\scriptstyle(R')}j_\phi{\scriptstyle(R')}dR'.
\end{align}

\noindent
The associated densities are obtained from the equation of state, i.e.:
\begin{equation}
{\textstyle \rho=\frac{m_iP(x,y)}{2k_BT(y)}}.    
\end{equation}

\subsection{Setup} \label{setup}

To perform a 2.5D simulation we assume that the FR and the CH have a symmetry in the $z$-direction. A FR  of characteristic length $L_0 \sim 500\,\textrm{Mm}$   \citep{Berger2014} is large enough to make the symmetry assumption  appropriate. Given this characteristic length $L_0$, the CHs result in areas ($A\sim w L_0$) between $7.5\times 10^4\,\text{Mm}^2$ and  $1.5\times 10^{5}\,\text{Mm}^2$ \citep{2017ApJ...835..268H,2019SoPh..294..144H}.

We select the same model parameters as in \citetalias{Sahade2020} (see their Sec. 2.5 for more details). We vary the CH parameters as in Table~\ref{tbl-ch} to obtain a set of 18 simulations, which are run with two different FRs to obtain more general results. The initial parameters describing FR1 and FR2 simulations are listed in Table~\ref{tbl-fr}.

\begin{figure} 
   \centering
   \includegraphics[width=0.4\textwidth]{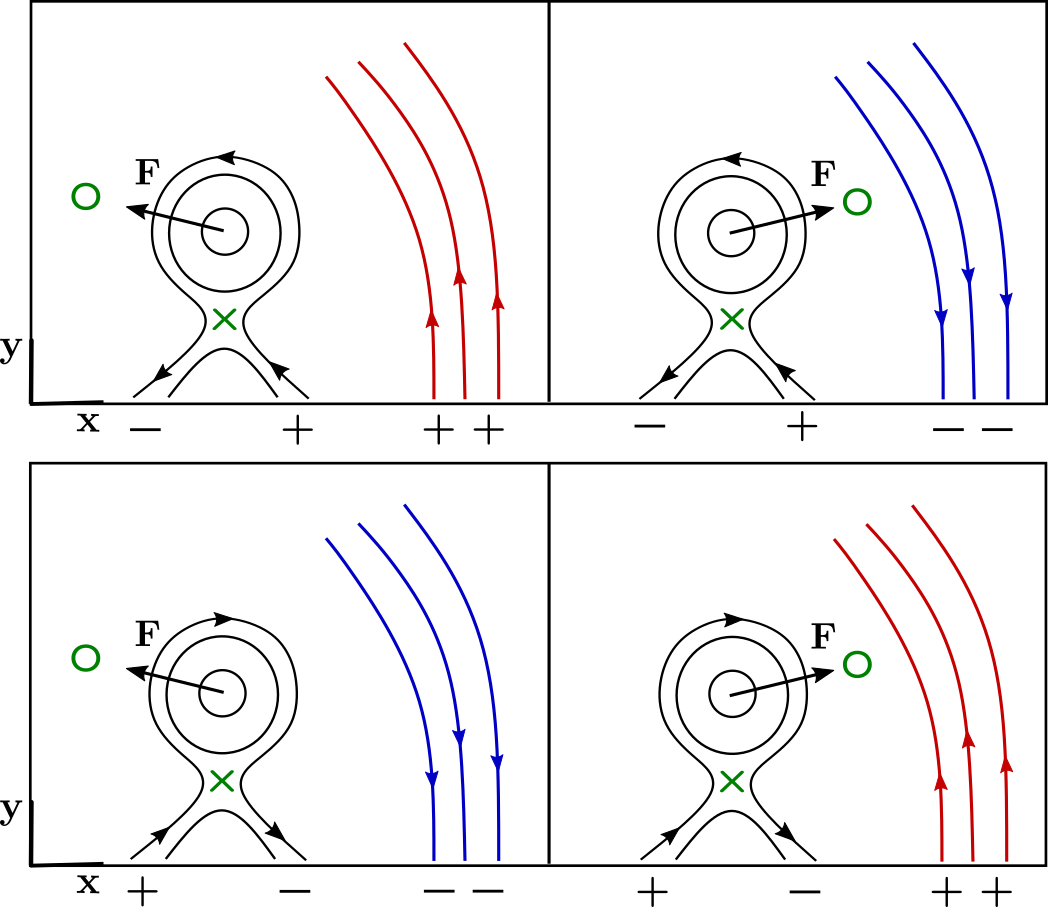}
      \caption{ Relative polarity scheme. The left panels show cases with aligned FR and CH polarities, for which a null point (green circle) forms on the opposite of the CH. Right panels show the anti-aligned configuration, for which the minimum forms between the FR and CH. The forces $F$ acting on the FR centre are directed towards the null point. The green $\times$ symbols indicate the site of reconnection. }
         \label{FigPol}
   \end{figure}

%
\begin{table}
\caption{CH parameters.}           
\label{tbl-ch}      
\centering                         
\begin{tabular}{c r c c}        
\hline\hline               
Case & $B_0\,$ & $D\,$& $W\,$ \\ 
  & (G) & (Mm)& (Mm) \\ 
\hline                        
    1 & $0.4$  & $150$ & $400$\\
    2 & $-0.4$  & $150$ & $400$\\
    3 & $0.8$  & $150$ & $400$ \\
    4 & $-0.8$  & $150$ & $400$ \\
    5 & $1.2$  & $150$ & $400$\\
    6 & $-1.2$  & $150$ & $400$\\
    7 & $1.6$  & $150$ & $400$\\
    8 & $-1.6$  & $150$ & $400$\\
    9 & $0.8$  & $250$ & $400$\\
    10 & $-0.8$  & $250$ & $400$\\
    11 & $0.8$  & $350$ & $400$\\
    12 & $-0.8$  & $350$ & $400$\\
    13 & $0.8$  & $150$ & $300$\\
    14 & $-0.8$  & $150$ & $300$\\
    15 & $0.8$  & $150$ & $500$ \\
    16 & $-0.8$  & $150$ & $500$ \\
    17 & $0.8$ & $150$ & $600$  \\
    18 & $-0.8$ & $150$ & $600$ \\
\hline                                   
\end{tabular}
\tablefoot{Parameter $B_0$ is the CH magnetic field strength, $D$ its distance to the FR, and $W$ is related to its width.}
\end{table}
%
\noindent
Cases in Table~\ref{tbl-ch} with positive (negative) magnetic strength $B_0$ are aligned (anti-aligned), meaning that the FR footpoint which is nearest to the CH has positive (positive) polarity, as shown in the left (right) top panel of Fig.~\ref{FigPol}. Note that the left panels of this figure are equivalent configurations (aligned), as are those on the right (anti-aligned). It is the relative polarity alignment what provides the different scenarios. The cases in the two upper panels represent the two possible relative FR-CH polarities.  

\begin{table}
\caption{Initial state parameters.}           
\label{tbl-fr}
\centering                         
\begin{tabular}{l c c}        
\hline\hline 

Parameter & \multicolumn{2}{c}{Value}\\
 & FR1 & FR2 \\
 
\hline
 $j_0\,(\text{stA}\,\text{cm}^{-2})$  & $435$& $295$\tablefootmark{a} \\
  $j_1\,(\text{stA}\,\text{cm}^{-2})$ & $455$& $300$ \\
  $T_{\text{\tiny{FR}}}\,(\text{MK})$& $1$& $4$\\
  $n_\mathrm{c}\,(\text{cm}^{-3})$ & $3\times10^8$& $4.5\times10^8$\\
    $n_\mathrm{CH}\,(\text{cm}^{-3})$      &\multicolumn2c{$2\times10^8$}  \\ 
  $h_0\,(\text{Mm})$      &\multicolumn2c{$30$}  \\ 
  $r\,(\text{Mm})$        &\multicolumn2c{$2.5$}\\
  $\Delta\,(\text{Mm})$        & \multicolumn2c{$0.25$} \\
  $d\,(\text{Mm})$   & \multicolumn2c{$3.125$} \\
  $M$        & \multicolumn2c{$1$}  \\
\hline
\end{tabular}
\tablefoot{Parameters $j_0$ and $j_1$ are the current densities inside the flux-rope in the $z$-direction and $\phi$-direction, respectively, $T_{\text{\tiny{FR}}}$ is the internal FR temperature, $n_\mathrm{c}$ is the numerical density at the base of the corona, $n_\mathrm{CH}$ is the numerical density for the CH, $h_0$ is the vertical position (height) of the FR, $r$ is its radius, and $\Delta$ is the thickness of the transition layer between the FR interior and the corona. Parameters $d$ and $M$ are the depth of the line dipole below the boundary surface and its relative intensity, respectively.\\
\tablefoottext{a}{ The value $j_0$ for FR2 was corrected because of a typo error in \citetalias{Sahade2020}.}}
\end{table}

\subsection{Results}
   \begin{figure} 
   \centering
   \includegraphics[width=0.49\textwidth]{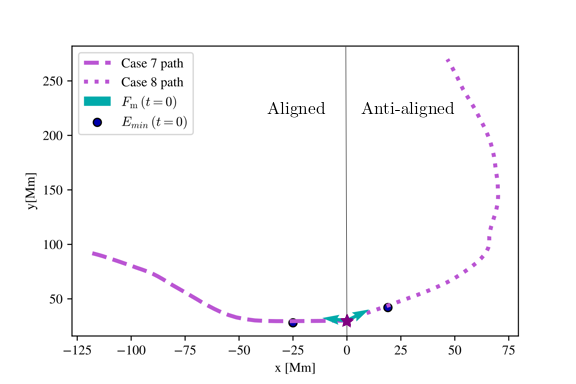}
      \caption{Paths of the FR centre as a function of time (dashed line: aligned case, dotted line: anti-aligned case). The purple star indicates the initial position of the FR centre, the blue circles the initial position of the magnetic null point ($E_{\textrm{min}}$) and the blue arrows the magnetic force direction on the FR ($F_{\textrm{m}}$) at $t=0\,$s.}
         \label{FigLorF}
   \end{figure}

As we mentioned in \citetalias{Sahade2020}, a minimum magnetic energy region (hereafter null point) arises at the position where the magnetic field of the CH and FR counteracts (green circles of Fig.~\ref{FigPol}). The FR-CH aligned polarity cases with positive $B_0$, have the null point to the left of the FR (negative $x$-positions, see top left panel of Figure \ref{FigPol}); while the FR-CH anti-aligned polarity cases with negative $B_0$, have the null point to the right of the FR (positive $x$-positions, see top right panel of Figure \ref{FigPol}). We note that the anti-aligned cases have the null point between the FR and the CH.  For both configurations, we calculate the forces acting on the FR centre at $t=0\,$s. We find that the magnetic force is the main contribution to the total force. The magnetic force of the simulated cases always points towards the magnetic null point, initially deflecting the FR towards this position. In the aligned cases the magnetic force points towards a direction away from the CH. The opposite occurs for the anti-aligned cases.

Figure~\ref{FigLorF} displays for FR2 the aligned case 7 (negative $x$ values) and the anti-aligned case 8 (positive $x$ values). In the figure we show the initial null point locations (blue circles), the initial magnetic forces on the FR centre (cyan arrows) and the resulting trajectories. The last point plotted corresponds to the time when the trajectory becomes parallel to the magnetic field: $700~\textrm{s}$ for case 7 and $1200~\textrm{s}$ for case 8. The trajectory of case 7 (dashed-line) deflects only once to the left, while that of case 8 (dotted-line) is first deflected to the right and later to the left, i.e. the FR suffers a double deflection. The general dynamics of the aligned cases are represented by case 7, while the anti-aligned cases are represented by case 8. It follows that the single or double deflection only depends on the relative alignment of the FR and CH polarities.

   \begin{figure} 
   \centering
   \includegraphics[width=0.45\textwidth]{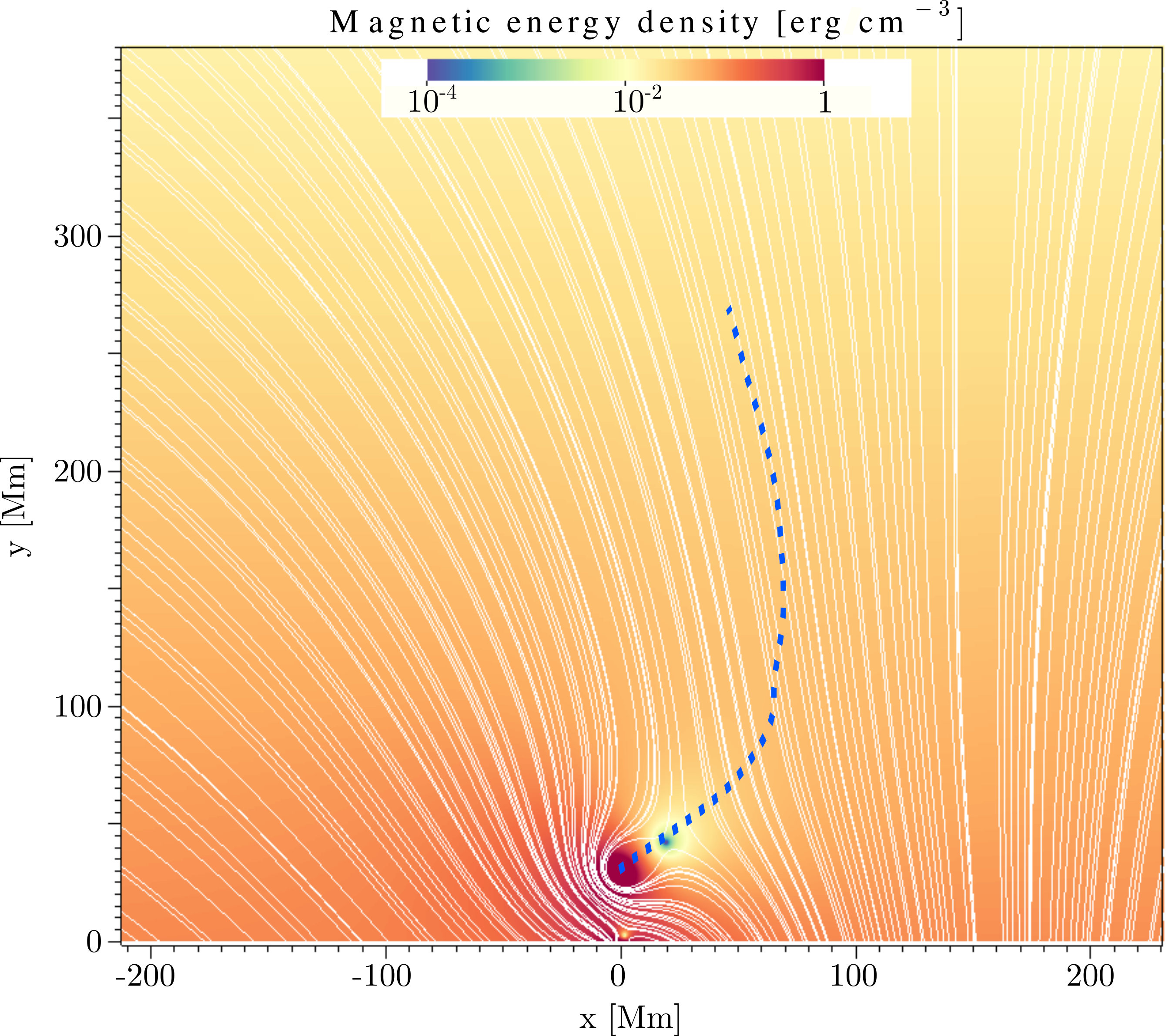}
      \caption{Initial magnetic energy distribution and magnetic field lines for case 8. The dotted line represents the path of the FR centre as a function of time until $t=1200~\textrm{s}$. The  animated evolution of this figure is available in the HTML version.      }
         \label{FigEmag}
   \end{figure}
To explain the double-deflection mechanism seen in anti-aligned cases, we analyse the evolution of the magnetic energy density, also considering the magnetic field lines. Figure~\ref{FigEmag} (see animated version available in the HTML version) displays the initial magnetic energy distribution and the magnetic field lines for case 8. The maximum energy (red in the colour scale) is located at the FR centre and the minimum (blue in the colour-scale) is located between the FR centre and the CH. The background energy decays with height, along the $y$-axis. Initially the FR travels towards the null point and collides with it. Then the null point stretches, leading to a low magnetic energy region that surrounds the FR front.
A turbulent zone is excited in this low magnetic energy region, whose plasma parameter $\beta\approx 1$,  entangling the magnetic field lines. The FR continues moving in the original direction, driving a coronal shock wave that compresses and bends the field lines.
This results in a magnetic force that eventually slows down the lateral motion of the FR, redirecting its trajectory to finally become parallel to the field lines. 
In summary, in the low corona and for an anti-aligned case, the magnetic force appears as the main driver of the FR trajectory evolution: it is firstly attracted toward the null point, then it surpasses the null point position due to inertia, until the magnetic force exerted by the compressed and bent CH magnetic field lines pushes the FR away. The final FR path is that of least resistance, i.e. it travels guided by the magnetic field lines towards a region of less magnetic energy. Thus, at later stages of evolution all FRs move away from the CH, independently of the polarity configuration, guided by the magnetic field.

   \begin{figure*}[ht!] 
   \centering
   \includegraphics[width=0.8\textwidth]{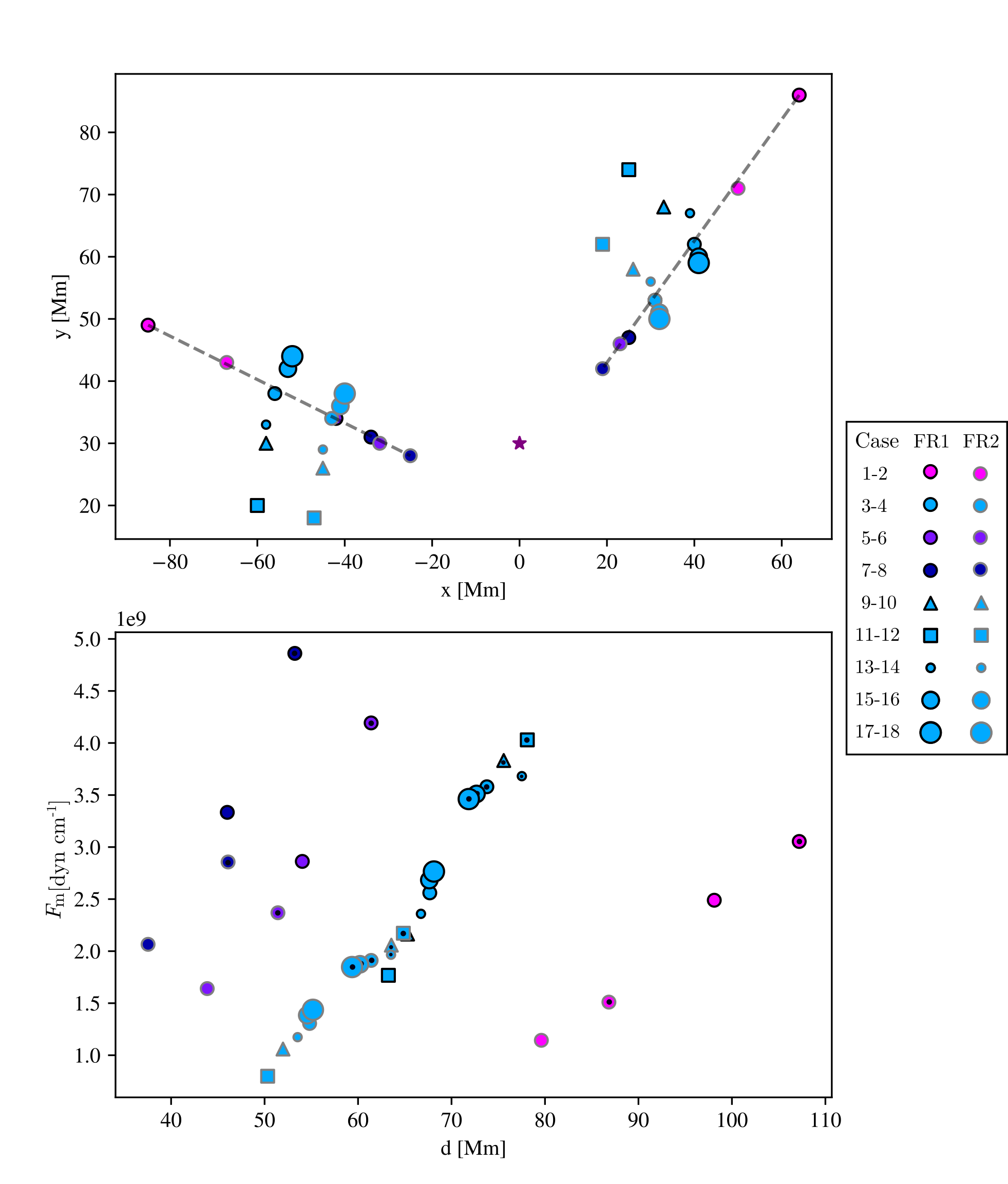}
   
      \caption{a) Null point locations at $t=0\,$s for all cases. The purple star represents the position of the FR, black edges correspond to FR1 simulations and grey edges to FR2 ones. Dashed lines indicate the location of null points for aligned (left) and anti-aligned (right) cases, with the CH at the same distance $D$ and same parameter $W$. The symbol colour changes with the strength $B_0$ of the magnetic field (from weaker to stronger: pink, light blue, blue violet and dark blue), the size changes with the parameter $W$ (bigger means wider), and the shape changes with the distance $D$ (from closer to farther: circle, triangle and square). b) Module of the magnetic force per unit length versus null point distance to the FR centre.  Symbols hold the same properties as in a). A black dot in the symbol centre indicates an anti-aligned case.    }
         \label{FigMin-FL}
   \end{figure*}

Figure~\ref{FigMin-FL}a shows the null point locations for all cases in Table~\ref{tbl-ch}. The purple star represents the location of the FR centre. The symbols with black and grey edges correspond to FR1 and FR2 simulation cases, respectively. The colour of the symbols changes with the strength of the magnetic field $B_0$: dark blue for stronger ($1.6~\textrm{G}$) and pink for weaker ($0.4~\textrm{G}$). The shape of the symbols change with the distance between the CH and the FR: circle ($D=150~\textrm{Mm}$), triangle ($D=250~\textrm{Mm}$) and square ($D=350~\textrm{Mm}$). Changes in  the CH parameter $W$ are represented with different symbol sizes: the smallest for $W=300~\textrm{Mm}$ and the largest for $W=600~\textrm{Mm}$. 

We note that for a given parameter $W$, at a given distance $D$, there is a linear correlation between  the location of the null point and the  magnetic field strength $B_0$. The larger the magnetic field strength, the nearest the null point is to the FR centre.
In Fig.~\ref{FigMin-FL}a the CH cases with  $W=400~\textrm{Mm}$ (same size symbols) and distance $D=150~\textrm{Mm}$ (circle symbols) generate the dashed grey lines of null point positions. The straight line with positive (negative) slope represents the cases with anti-aligned (aligned) polarity. 

Also, for a given distance $D$ (same symbol shape) and a given magnetic field strength $B_0$ (same colour), a change in $W$ (different sized circles) yields a curve that crosses the straight lines. For the aligned polarity cases the larger the CH width, the higher the null point position. Conversely, for the anti-aligned cases the larger the CH width, the lower the height of the null point. 

Moreover, for a given magnetic field strength $B_0$ (same symbol colour) and a given parameter $W$ (same symbol size) a change in the distance (different symbols) yields a curve that crosses the straight line.  
For the aligned polarity cases the larger the distance, the lower the height of the null point. On the contrary, for the anti-aligned cases the larger the distance, the higher the null point location. 

In Fig.~\ref{FigMin-FL}b a positive correlation between the magnetic force magnitude per unit length and the null point distance ($\textrm{d}$) to the FR can be noticed. This correlation relies on the magnetic strength $B_0$, resulting in four different slopes (linking both polarities and FR configurations); the larger the magnetic field strength, the larger the slope. The anti-aligned cases are differentiated from the aligned ones by a black dot in the symbol centre. 

This analysis sheds light on the relationships between null point location, distance between CH and FR, strength of the CH magnetic field and CH area. A deeper study that considers a wider range of parameters is required, so as to predict null point locations more precisely.

\section{An observational case  study}
 \begin{figure} 
   \centering
   \includegraphics[width=0.4\textwidth]{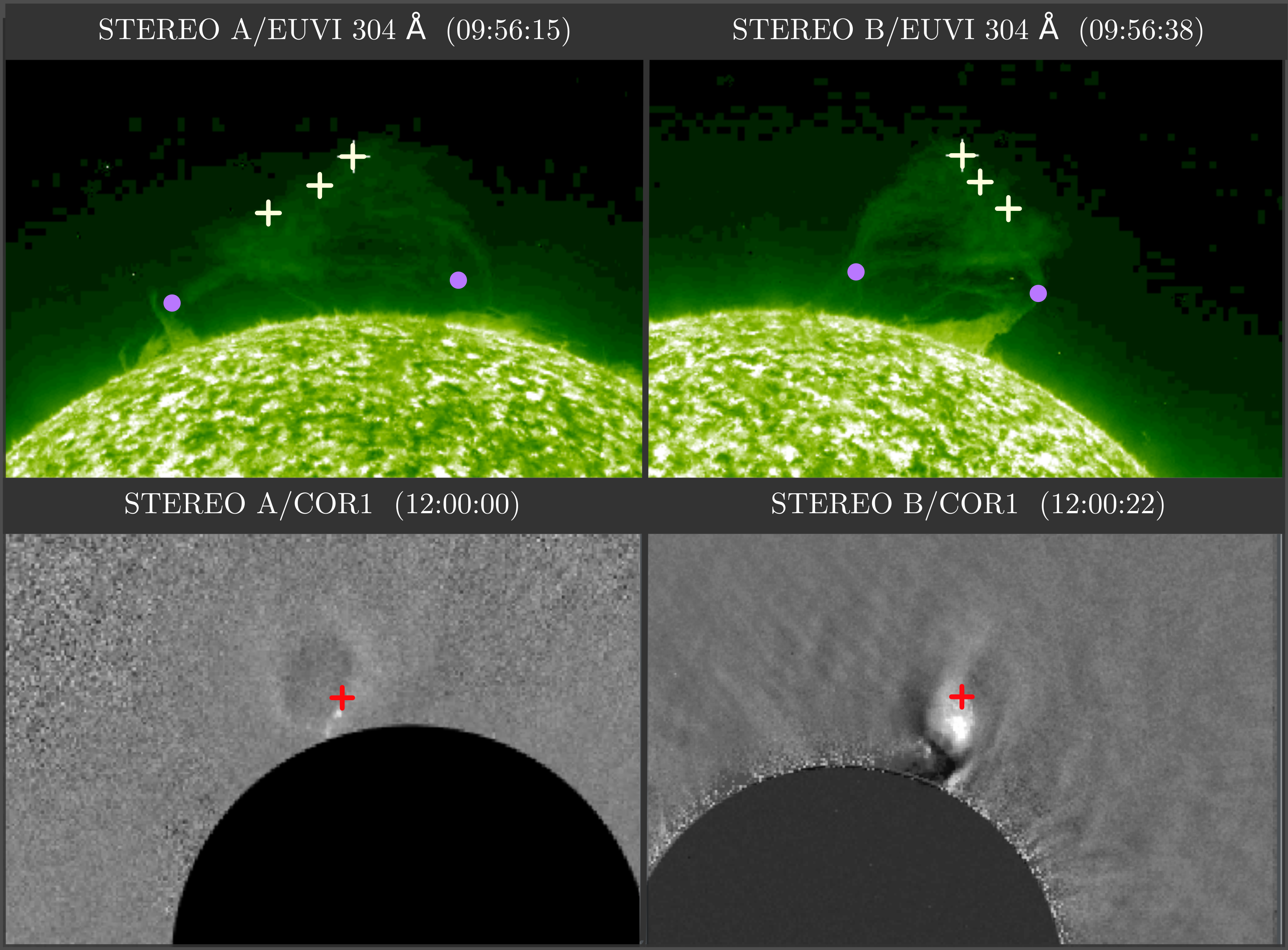}
      \caption{Tie-pointing reconstruction of the 30 April 2012 prominence. 
      The three white plus symbols in the top panels mark the wide apex considered in STEREO EUV images. The red plus symbol in the bottom panels show the apex in the STEREO coronagraph images. 
              }
         \label{FigTP}
   \end{figure}

   \begin{figure} 
   \centering
   \includegraphics[width=0.4\textwidth]{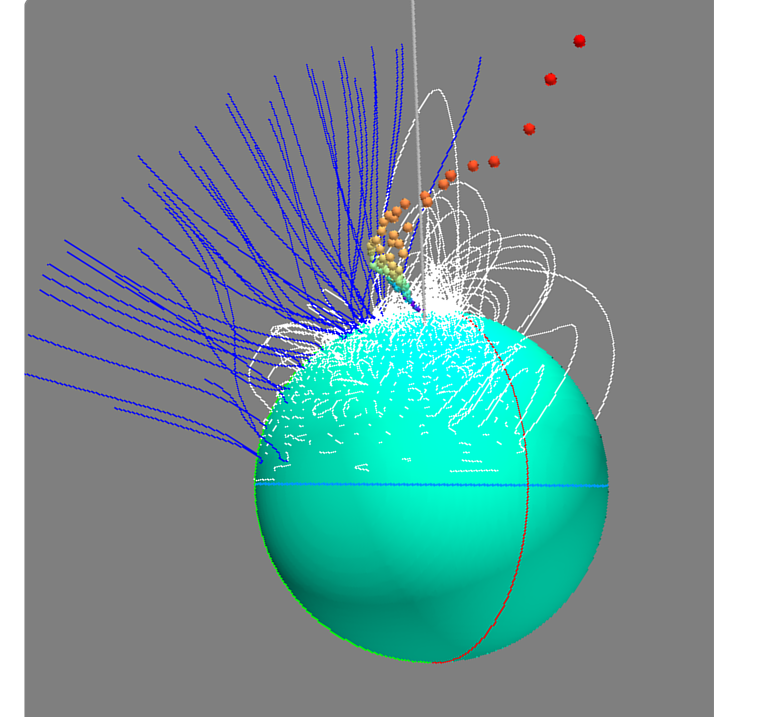}
      \caption{3D evolution of the prominence apex since 04:06 UT to 13:30 UT, from bluish to reddish spheres. The grey straight line represents the radial direction at the initial position of the apex. Blue lines belong to the open field lines of the CH and the white ones are the closed magnetic field lines. Red meridian points at the Earth direction, green meridian at the STEREO B direction, and the light-blue line marks the solar Equator.
              }
         \label{Figtray}
   \end{figure}

\subsection{Methods and techniques}
To compare our model with observations, we analyse the trajectory of an eruptive prominence associated with a CME that occurred on April 30, 2012. During the outward motion of the prominence, the trajectory changed its direction twice, i.e. the eruption suffered a double deflection. A nearby CH, together with the eruptive prominence, constitute a relatively isolated scenario that can be approximated by one of the schemes in Fig.~\ref{FigPol}. 
The eruption occurred to the east of the northern hemisphere starting at 04:06 UT. To reconstruct the 3D trajectory of the eruption we track the prominence cold material by means of EUV images from the Extreme-Ultraviolet Imager (EUVI) of the SECCHI instrument suite (Sun-Earth Connection Coronal and Heliospheric Investigation, \citealt{Howard2008}) and white-light images from the SECCHI COR1 coronagraphs on board the STEREO A (ST-A) and STEREO B (ST-B) spacecraft. From 04:06 UT to 11:56 UT we use EUVI $304\,$\r{A} images from ST-A and ST-B. Running difference images from COR1 on board ST-A and ST-B are used to track the eruption at a later stage, from 11:45 UT to 13:30 UT. 

The position of the prominence in time is determined by means of the tie-pointing reconstruction technique \citep[hereafter triangulation,][]{Inhester2006, Mierla2010}, which is applied to the apex, chosen to represent the global motion of the ejection. To allow for a better description of the trajectory, we consider a wide apex by triangulating three points from the prominence front at each analyzed time (see plus symbols in Figure \ref{FigTP}, top panels). 
The sequences of the three points follow the same trend, therefore, we consider that they represent well the global motion of the ejection. We tracked other points of the prominence and we found that they also follow the same trend. In addition, we triangulate the prominence feet positions to estimate their location and length. Figure~\ref{FigTP} upper panels show, in addition to the points characterising the apex (white plus symbols), those representing the prominence feet (pink dot symbols) in the $304\,$\r{A} filter of EUVI-A and EUVI-B; while the lower panels of the figure show the prominence apex in COR1-A and COR1-B.

To analyse the magnetic structure surrounding the ejection area we use the Potential Field Source Surface (PFSS) model by \citet[][]{2003SoPh..212..165S}.

\subsection{Prominence evolution: trajectory and magnetic environment}

\begin{figure}[ht!] 
   \centering
  
   \includegraphics[width=0.45\textwidth]{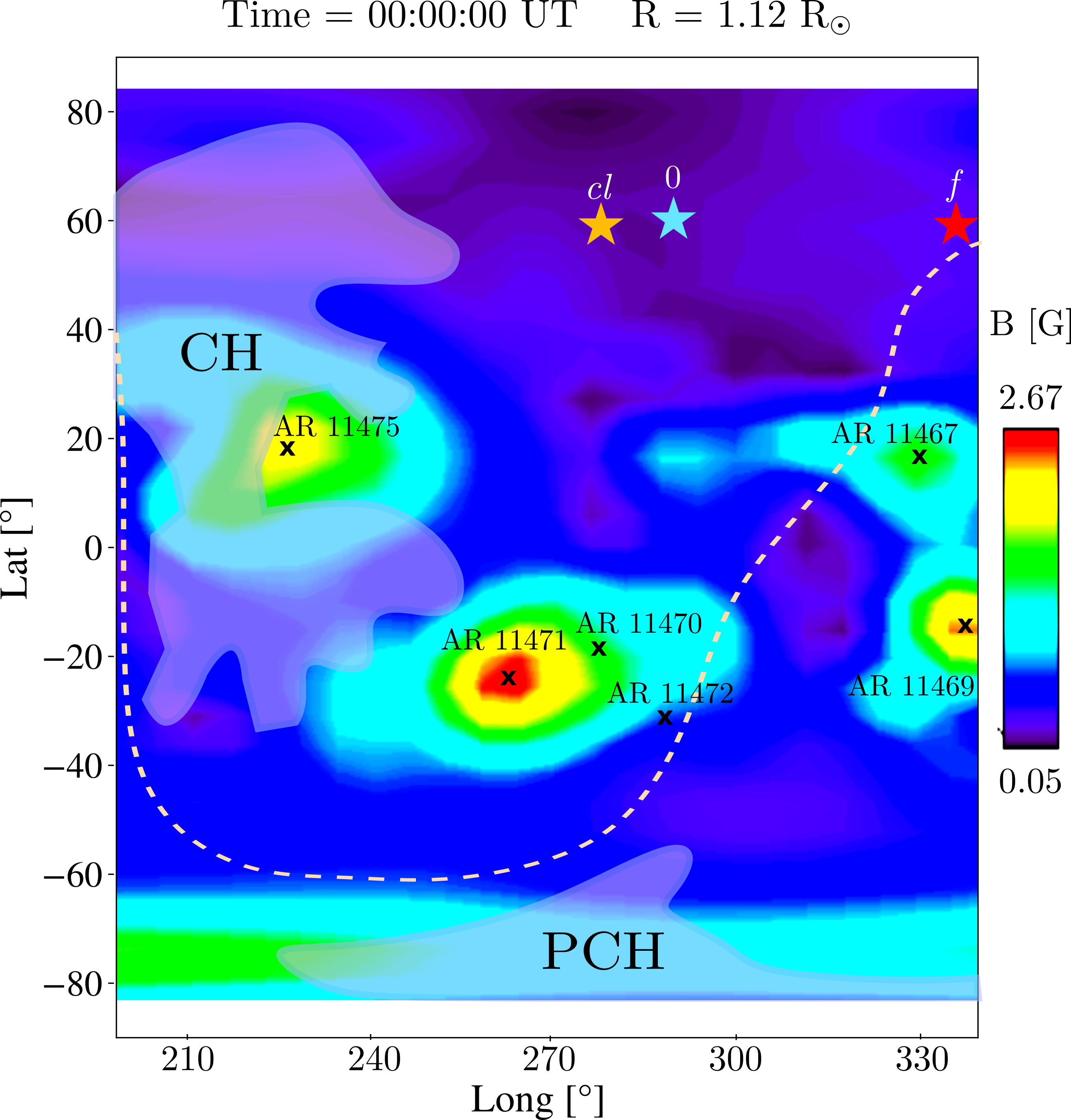}
      \caption{Magnetic strength map at $R=1.12\, R_{\odot}$ (initial height of the prominence), from longitude $200^{\circ}$ to $340^{\circ}$, on 30 April 2012 at 00:00 UT. The stars represent the initial (cyan), the closest to CH (orange), and the last measured (red) prominence position. The CHs, the active regions (NOAA) and the heliospheric current sheet (WSO) are drawn over the map. PCH stands for polar coronal hole. The map has been provided by the Community Coordinated Modeling Center at Goddard Space Flight Center through their public Runs on Request system (http://ccmc.gsfc.nasa.gov) with the PFSS model. 
              }
         \label{FigPfss}
   \end{figure}

The 3D trajectory determined by applying the triangulation method is displayed in Fig.~\ref{Figtray}. As can be seen from the figure, the eruption initially approaches the negative open magnetic field lines (blue lines, reconstructed from PFSS) and then moves away from it, following a new direction. The open magnetic field lines correspond to a CH that was catalogued as SPoCA 4548 \citep{SPoCA1,SPoCA2} when it was facing Earth on 4 April 2012. 
We measure the northeast and southwest feet of the erupting prominence on 30 April 2012 at $t=05:06~\textrm{UT}$ and at a height $R=1.09\,R_{\odot}$, to be located at $(67^{\circ},264^{\circ})$ and $(60^{\circ},307^{\circ})$ respectively. The initial measured height, average latitude and Carrington longitude of the prominence are $(R_0,\theta_0,\phi_0)=(1.12\,R_{\odot},61.6^{\circ},289.4^{\circ})\,$ at 04:06 UT (cyan point in Fig.~\ref{Figtray} and cyan star in Fig. \ref{FigPfss}). Then, the prominence apex moves towards the closest minimum magnetic energy region, approaching the CH until $(R_{cl},\theta_{cl},\phi_{cl})=(1.45\,R_{\odot},59.4^{\circ},278.1^{\circ})\,$ at 10:46 UT (orange points in Fig.~\ref{Figtray}). Next, the prominence apex deflects away from the CH towards the heliospheric current sheet and the final measured position is $(R_f,\theta_f,\phi_f)=(2.67\,R_{\odot},61.4^{\circ},336.8^{\circ})\,$ at 13:30 UT (higher red point in Fig.~\ref{Figtray} and red star in Fig.~\ref{FigPfss}). Thus, the prominence suffers a double deflection: first it moves towards lower longitudes when approaching the CH (from $289^{\circ}$ to $278^{\circ}$) and afterwards it moves away from it, increasing its longitude from $278^{\circ}$ to $336^{\circ}$ and reaching the heliospheric current sheet. 

The prominence erupts under the scenario described in Figure~\ref{FigPfss}. The figure shows a map of the magnetic field strength between 200$^\circ$ and 340$^\circ$, where we indicate the magnetic structures as catalogued by the National Oceanic and Atmospheric Administration (NOAA) and Wilcox Solar Observatory (WSO) Source Surface Synoptic Charts.
The pink shaded areas enclose the CHs, with size and location estimated from the open lines of the PFSS and the dark regions of the ST-B/EUVI $195\,$\r{A} filter. The cyan star indicates the source region (SR), the orange star indicates the closest position to the CH, and the red star shows the last position measured. Several active regions can be noticed in the map. Note that, as a result of the projection, AR 11467 may appear to be at a similar distance from the SR as the CH to the northeast. But in fact, the nearest region of that AR (namely its eastern extension in the cyan colour) is almost two times farther away than the nearest edge of the CH.
The dashed line represents the heliospheric current sheet. Between the CH and the SR there is a zone of low magnetic strength (between the cyan and orange star). Figure~\ref{FigPfss} also shows a region of even lower magnetic strength towards the north and the southwest of the SR. 

\begin{figure}[ht!] 
   \centering
  
   \includegraphics[width=0.45\textwidth]{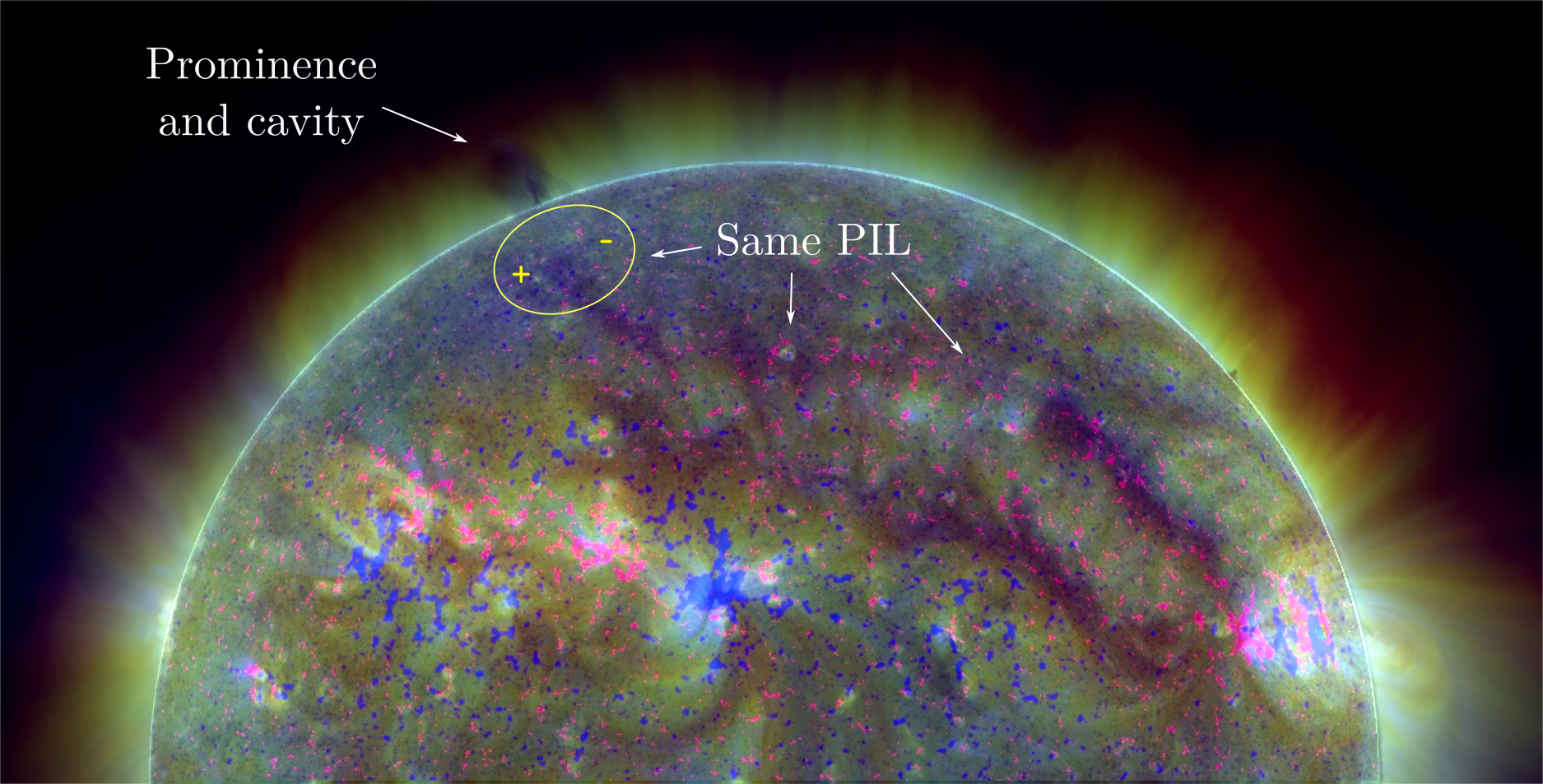}
      \caption{SD0/AIA composite image of $211\,$\r{A}, $193\,$\r{A} and $171\,$\r{A} at 01:26 UT overlapped to an SDO/HMI image at 01:30 UT on the day of the event. In the latter, magenta corresponds to magnetic fields with positive polarity and blue to negative ones. The prominence and its overlying cavity, as well as the polarity inversion line (PIL) were the prominence presumably lies are indicated by arrows. The white circle denotes the region were post-eruptive loops form at a later time after eruption.
              }
         \label{FigComposite}
   \end{figure}

As the eruption took place at very high latitudes ($\sim\,60^{\circ}$) and close to the limb from Earth's perspective, the characterisation of the photospheric magnetic field was not straightforward. The identification of the polarities at both sides of the neutral line was based on the following reasoning, arising from Figure \ref{FigComposite}. The figure depicts SDO/AIA coronal wavelengths overlapped to an SDO/HMI magnetogram, both at a time before eruption of the prominence. The prominence, surrounded by a circular coronal cavity, is seen above the north-northeast limb. At a later time, well after the eruption (starting ca. 12:00 UT), post eruptive loops can be discerned particularly in SDO/AIA 211\,\r{A}. These loops are not seen at the time of Figure \ref{FigComposite}, but the approximate region where they form is encircled in the figure. In fact, the loops form above the polarity inversion line seen as a dark channel ending at the prominence location, close to the limb. Therefore, it appears highly plausible that the eruptive prominence was sitting at that same neutral line, which enables identification of the magnetic field polarity at its sides. According to the information provided by the SDO/HMI magnetogram, there is predominance of negative polarity to the northwest of the neutral line holding the prominence, and positive polarity to the southeast. This scenario, together with the fact that the magnetic field polarity of the CH is negative, suggests that this event is an example of an anti-aligned case.

\subsection{Simulated event}
\begin{figure} 
    \centering
    \includegraphics[width=0.48\textwidth]{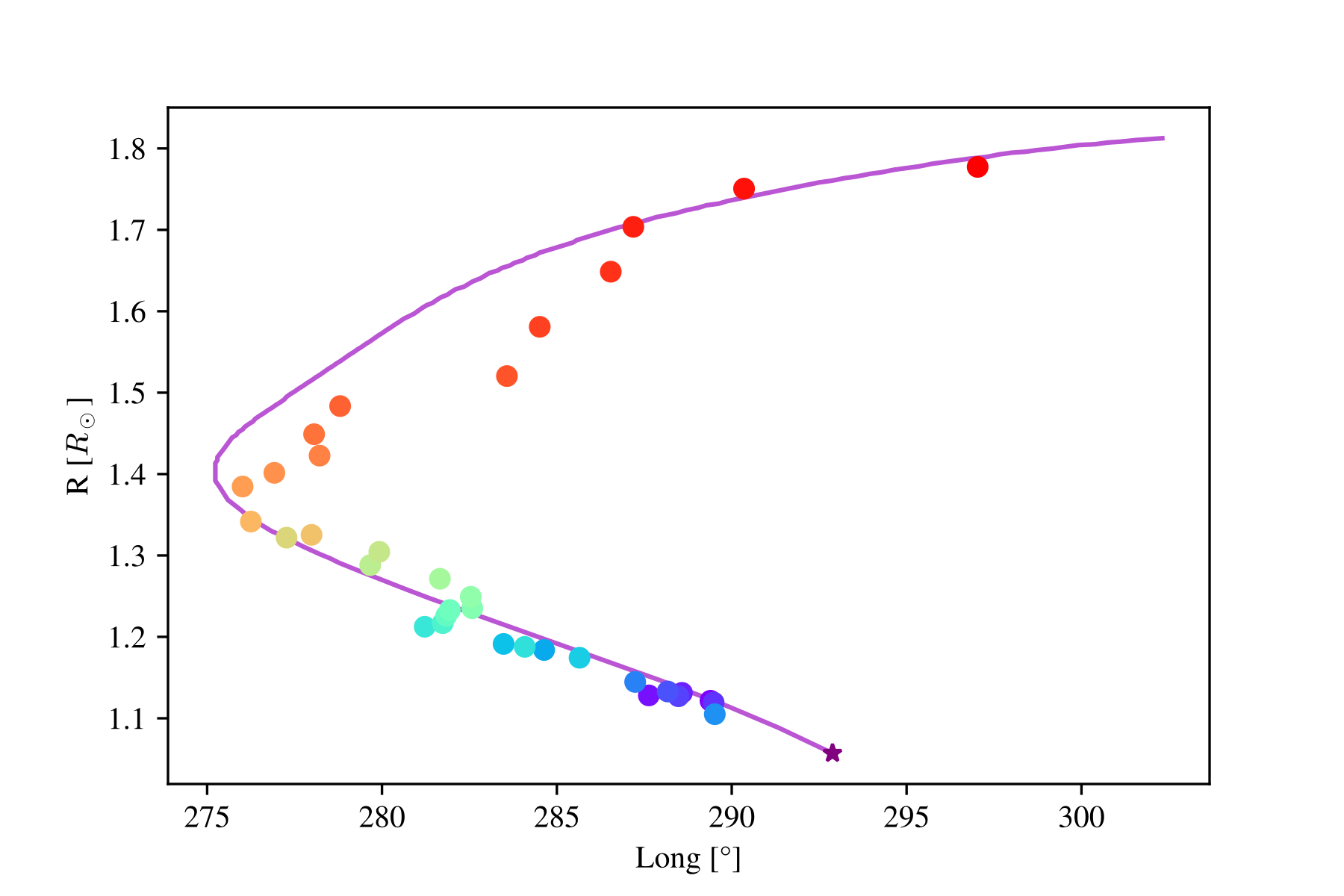}
    \caption{Height versus average longitudinal deflection of the 30 April 2012 prominence (dots) and the trajectory of the simulated FR (solid line) with the purple star showing the initial FR centre position.}
    \label{FigObs-Simu}
\end{figure}

\begin{table}
\caption{Parameters used for the simulation of the observed event.}           
\label{tbl-observ}
\centering                         
\begin{tabular}{l c }        
\hline\hline 
Parameter & Values\\
\hline
 $j_0\,(\text{stA}\,\text{cm}^{-2})$  & $435$ \\
  $j_1\,(\text{stA}\,\text{cm}^{-2})$ & $455$ \\
  $T_{\text{\tiny{FR}}}\,(\text{MK})$& $0.8$\\
  $n_\mathrm{c}\,(\text{cm}^{-3})$ & $3\times10^8$\\
  $n_\mathrm{CH}\,(\text{cm}^{-3})$      &$2\times10^8$  \\ 
  $h_0\,(\text{Mm})$      &$40$  \\ 
  $r\,(\text{Mm})$        &$2.5$\\
  $\Delta\,(\text{Mm})$        & $0.25$ \\
  $d\,(\text{Mm})$   & $3.125$ \\
  $M$        & $1$  \\
  $B_0\,(\text{G})$ & $-0.6$\\
  $D\,(\text{Mm})$ & $300$\\
  $W\,(\text{Mm})$ & $400$\\
\hline
\end{tabular}
\tablefoot{Parameters $j_0$ and $j_1$ are the current densities inside the flux-rope in $z$-direction and in $\phi$-direction, respectively, $T_{\text{\tiny{FR}}}$ is the internal FR temperature, $n_\mathrm{c}$ the numerical density at the base of the corona, $n_\mathrm{CH}$ the numerical density for the CH, $h_0$ the vertical location (height) of the FR, $r$ its radius, and $\Delta$ the thickness of the transition layer between the FR interior and the corona. Parameters $d$ and $M$ are the depth of the line dipole below the boundary surface and its relative intensity, respectively. The CH parameters are $B_0\,$, $D\,$ and $W$.
}
\end{table}
\begin{figure*} 
    \centering
    \includegraphics[width=0.95\textwidth]{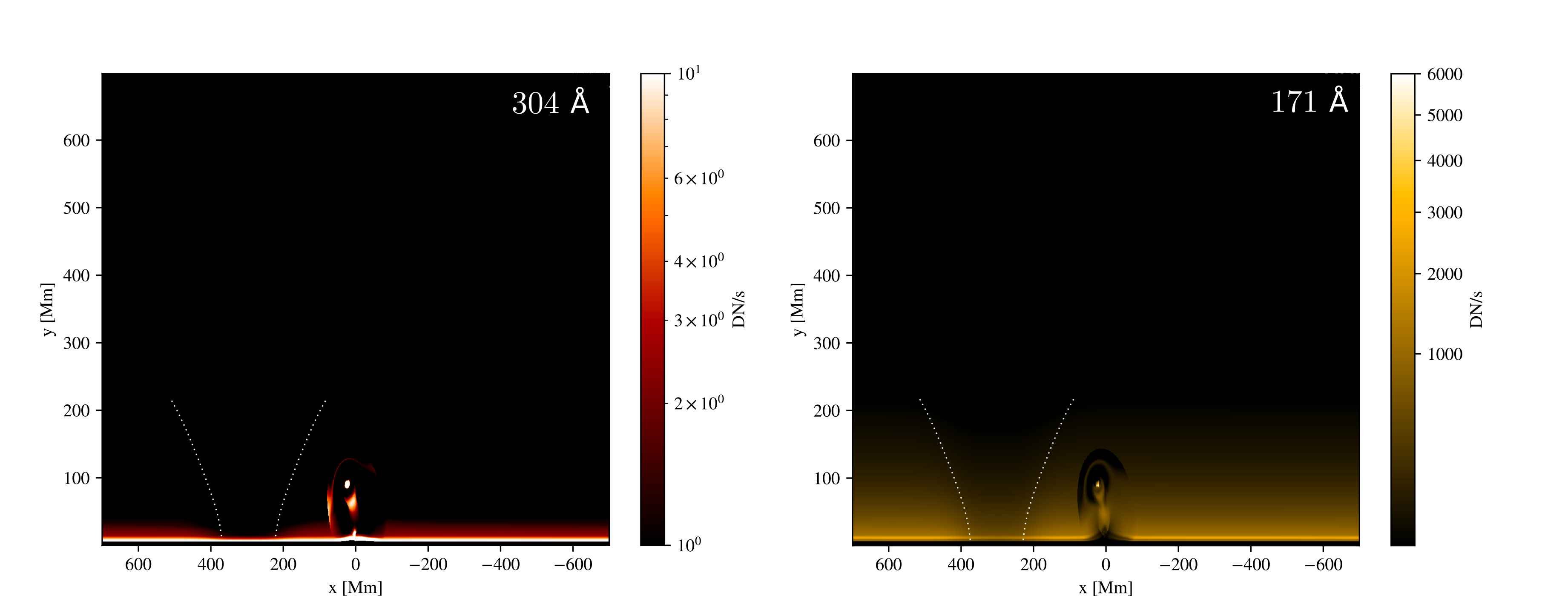}
    \caption{Synthetic EUV emission for filters $304\,$\r{A} and $171\,$\r{A} using the AIA response temperature. White dotted-lines represent the estimated boundary of the CH. The animation of the full evolution is available at the HTML version.}
    \label{FigEmision}
\end{figure*}

From the observational event analysed in the previous section, a question arises: is it possible that the mere presence of the CH results in the attraction of the FR structure? The numerical study carried out here together with information on the magnetic field configuration suggests that this may be the case if the FR and CH are magnetically anti-aligned. 

Considering that the latitudinal displacement is negligible in comparison to the longitudinal one, we perform a simulation to emulate the deflection in the latter direction. The axis of the prominence is oriented along the northeast–southwest direction and its length is $\sim 300~\textrm{Mm}$, which is large enough to assume a 2.5D approximation. Our model simulates the transverse motion projected on the longitudinal coordinate. 
Table \ref{tbl-observ} lists the parameters used for the simulation. We set the temperature $T_\mathrm{FR}$ to emulate the emission of the prominence in the corresponding EUV filters. The CH parameters are estimated from the magnetic maps ($B_0$) and from the EUVI-B $195\,$\r{A} filter ($D$ and $W$).

Figure \ref{FigObs-Simu} shows the average longitudinal deflection for the triangulated event (coloured dots), and the deflection of the simulated case (solid line). We note that a model of anti-aligned polarity is able to reproduce the deflection of the observed event. Taking into account that many effects are not considered in this simple model, the match between the simulation and the observation suggests that the proposed magnetic configuration is a possible cause of the prominence approach to the CH.

Figure \ref{FigEmision} shows the synthetic emission in the $304\,$\r{A} and $171\,$\r{A} wavelengths generated by the FoMo tool \citep{VanDoorsselaere2016} at $t=200\,$s. The code sets the thickness of simulation to 1\,Mm since it is 2.5D. The subdense CH is set to the left of the FR structure, while the simulated FR is seen to emit in both wavelengths. The full evolution can be seen as an animation in the HTML version. During its evolution a subdense cavity is formed around the FR (clearly noticed in the 171~\AA~ filter). The movie also shows the formation of waves and shocks.

\section{Conclusions}

Determining the mechanisms and the scenarios that produce deflections during the evolution of CMEs is a crucial step for improving space weather predictions. Many studies suggest that open magnetic fluxes behave as ``magnetic walls'' that push CMEs away \citep[e.g.,][]{Cremades2006,Gopalswamy2009,Gui2011,Cecere2020}. 
However, there are few reported events where the eruptions first approach open magnetic fluxes and then propagate away from them \citep{Jiang2007ApJ,Yang2018,Sieyra2020}. In these observational cases the CME structure is close to, not only a CH, but also other magnetic structures, such as an active region or a pseudostreamer, making it difficult to distinguish which of all the structures are responsible for the double deflection and to what extent. Therefore, it is important to study cases in which the FR-CH system is almost isolated, both observationally and numerically.

In our previous work \citepalias{Sahade2020} we studied the FR-CH interaction numerically, simulating cases with aligned polarities between the two structures, and showing the importance of the null magnetic region to predict the FR early evolution. 
Here, we used the same 2.5D model for which the FR-CH interaction could be studied in isolation. Through numerical simulations we found that the FRs evolve towards the  null magnetic region whose position is determined by the FR and CH parameters. Particularly, the site where the null region is formed depends on the relative polarity alignment between FR and CH. During the early evolution of the system, the null point attracts the FR, in the aligned cases deflecting it away from the CH and in the anti-aligned cases causing that the FR approaches the CH. However, as in many observational cases mentioned above, in their later evolution all FRs  move away from the CH, guided by the magnetic field, independently of the relative FR-CH polarity alignment. The physical mechanism is the same for both alignment cases, but the phenomenological result is different: in the low corona, the anti-aligned cases will exhibit a double-deflection trajectory, while the aligned cases will display a single deflection trajectory. 

Performing several numerical simulations of different cases, varying FR properties and CH parameters, we found: 

\begin{enumerate}
    \item  For a given CH width, at a given distance to the FR, there is a linear correlation between the position of the null point and the magnetic field strength. The larger the magnetic field strength, the nearest the null point to the FR centre.
    \item In the aligned (anti-aligned) cases, for a given distance and a given magnetic field strength, the larger the  width the higher (lower) the null point location.
    \item In the aligned (anti-aligned) cases, for a given magnetic field strength and a given width, the larger the distance the lower (higher) the null point location.
    \item There is a positive correlation between the magnetic force magnitude per unit length and the null point distance to the FR. This correlation depends on the magnetic field strength: the larger the magnetic field strength, the larger the slope of the correlation.
\end{enumerate}

To compare our model with observations we analyse an event occurred on 30 April 2012. We consider the system as a relatively magnetically isolated scenario, where an ejected prominence is attracted by a nearby CH before it is deflected away from it.  
A careful analysis of AIA low coronal wavelengths, together with HMI photospheric magnetic fields (see Fig.~\ref{FigComposite}) reveals the magnetic polarities at the sides of the eruptive prominence. The analysis indicates that the magnetic polarities of the prominence source and of the CH are configured in an anti-aligned fashion. 
Assuming this particular alignment between both structures in a simple numerical model (where many effects were not considered) and approximate parameters estimated from the observations, we were able to reproduce the measured longitudinal deflection of the analysed event. This supports the conjecture that the presence of the CH and the magnetic topology that it produces are responsible for the double deflection of the event.
We can thus suggest that the null point first attracts the prominence towards the CH until the structure of the open magnetic fields deflects it away. Finally, the prominence follows the least resistance path towards the heliospheric current sheet.

Although, as shown in the literature, the scenarios in which the polarities are anti-aligned (see Sec.~\ref{setup}) seem to be common, most studies show that CMEs are deflected away from  CHs. However, these facts are not necessarily incompatible. On one hand, most of the observational events analysed in the literature are restricted to the high coronal region, so that a CME could have approached a CH at an early stage and thus, the double deflection would have not been reported. On the other hand, in studies that include the low corona, the deflection toward a CH may have been interpreted as the action of other magnetic structures pulling the CME towards the CH, without considering the mere presence of the CH as the possible cause of the attraction.
Some questions arise from this new insight: How common is the anti-aligned configuration? Is there a relation between the  deflection rate and the FR-CH alignment? To answer these questions, more FR-CH interactions should be studied in detail at low coronal heights.

\begin{acknowledgements}
We thank the reviewer, Dr. Paulett Liewer, for her valuable suggestions that helped us to substantially improve the previous version of this manuscript.
 AS is doctoral fellow of CONICET. MC, AC and HC are members of the Carrera del Investigador Cient\'ifico (CONICET). AS, MC and AC acknowledge support from ANPCyT under grant number PICT No. 2016-2480. AS and MC also acknowledge support by SECYT-UNC grant number PC No. 33620180101147CB. HC acknowledges support from UTN grant UTI4915TC. Also, we thank the Centro de C\'omputo de Alto Desempe\~no (UNC), where the simulations were carried out. Authors acknowledge use of data from STEREO (NASA) produced by the SECCHI consortium.
\end{acknowledgements}

\bibliographystyle{aa} 
\bibliography{biblio} 

\end{document}